\documentstyle[12pt,graphicx]{article}
\title
{\bf Half-quantum vortices in strongly correlated Bose liquids.}
\author{G.E. Volovik
\\Low Temperature Laboratory, Helsinki University of
Technology\\
P.O.Box 2200, FIN-02015 HUT, Finland\\
and\\
  L.D. Landau Institute for
Theoretical Physics\\  Kosygin Str. 2, 117940 Moscow, Russia}
\begin{document}
\maketitle

\abstract{
We discuss the structure of a vortex  in a superfluid Bose
liquid with a suppressed Bose-Einstein condensate and an intensive pair
correlated condensate.  The vortex represents the pair of half-quantum vortices
topologically confined by the soliton.}

\vspace{10mm}

In strongly interacting and strongly correlated Bose liquid -- superfluid
$^4$He -- the density of the Bose-condensate is small compared to the total
mass density of the liquid, $\rho_0\ll \rho$. This fact leads to
speculations that the rest part of the liquid can be described in terms of
the Cooper-like pair-correlated condensate (the most recent discussion of
this idea is in Ref. \cite{Pashitskii}; references to the older
papers can be found in
\cite{NepomnyashchiiPashitskii}). We would like to point out that if 
this idea is
correct, it must have a pronounced effect on the core structure of the
elementary vortex, which must be  non-axisymmetric.

Let us accept that the Cooper-like pair condensate is dominating and gives the
dominating contribution to the superfluid density. Then the phenomenological
free energy describing interacting pair- and Bose- condensates, which contains
all the relevant physics, can be written as follows
\begin{equation}
F={1\over 2}\rho {\bf v}_{\rm s}^2 + \epsilon(\rho) + F\{\Psi\} ~.
\label{F}
\end{equation}
Here ${\bf v}_{\rm s}$
\begin{equation}
{\bf v}_{\rm s} ={\hbar\over 2m}\nabla \phi ~,
\label{SuperfluidVelocity}
\end{equation}
is the superfluid velocity expressed in terms of the of pair condensate phase
$\phi$; and $\rho$ is the total mass density of the liquid (we consider
$T\rightarrow 0$). The single-particle Bose-condensate produces the small
correction which as we assume can be described in terms of the Ginzburg-Landau
functional, the last term in Eq.(\ref{F}).  The Galilean invariant
Ginzburg-Landau functional for the wave function of Bose condensate
$\Psi=|\Psi|e^{i\Phi}$ has the form
  \begin{equation}
F\{\Psi\}= {\beta\over 2\rho}\left(|\Psi|^2-\rho_0\right)^2+
  {\hbar^2\over 2m^2}\left|\left(-i\nabla -m{\bf v}_{\rm
s}\right)\Psi\right|^2 + \tilde\alpha |\Psi|^2 \sin^2\left(\Phi-{\phi\over
2}\right) .
\label{FGL}
\end{equation}
Here $\rho_0\ll
\rho$  is the density of the Bose
condensate in equilibrium, which is much smaller that the total density. The
last term in Eq.(\ref{FGL}) is the Josephson interaction coupling the two
condensates; it provides the phase coherence of the two condensates in
equilibrium, when $\Phi={\phi\over 2}$ and thus the two condensates have the
common superfluid velocity.

\begin{figure}
\centerline{\includegraphics[width=0.8\linewidth]{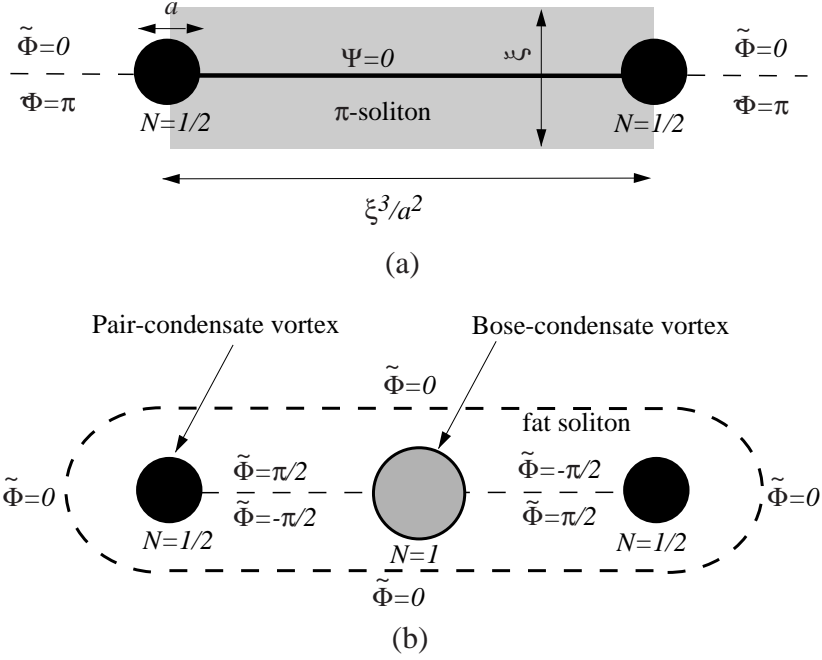}}
   \caption{Asymmetric vortex as pair of half-quantum vortices. (a) Vortex
structure in case of strong pinning of the Bose-condensate phase $\Phi$ by the
phase
$\phi$ of the pair condensate. Half-quantum vortices of the pair condensate are
confined by
$\pi$-soliton of the Bose-condensate. Within the soliton  the Bose-condensate
order parameter crosses zero.  (b) Weak-pinning case. The amplitude 
of the Bose-condensate
order parameter has equilibrium value
$\sqrt{\rho_0}$ everywhere except for the core of the $N=1$ vortex in the
Bose-condensate. In both Figures thin dashed lines terminating on half-quantum
vortices are lines where the phase $\tilde\Phi$ has a $\pi$-jump 
required by the
Aharonov-Bohm effect experienced by the Bose condensate in the presence of the
half-quantum vortices of pair condensate.}
   \label{ConfinementHalfQFig}
\end{figure}

We are interested in the structure of the $N=1$ vortex in this mixture of
two condensates. Around the
$N=1$ vortex the phase of the Bose-condensate $\Phi$ changes by
$2\pi$, while the phase $\phi$ of the pair-condensate wave-function changes by
$4\pi$. In other words, from the point of view of the pair condensate, the
$N=1$ vortex is doubly quantized. If $\rho_0=0$, i.e. in case of pure pair
condensate, the elementary vortices of pair condensate have $2\pi$ winding of
$\phi$, and thus they have twice smaller elementary circulation
$\kappa_2={1\over 2}\kappa_0$, where
$\kappa_0=2\pi\hbar/m$. From the point of view of the Bose condensate they
represent vortices with $N=1/2$ (discussion of the  half-quantum
vortices -- Alice strings -- can be found in the book \cite{VolovikBook}). For
nonzero but small
$\rho_0\ll
\rho$,  the half-quantum
vortices are combined into pairs forming $N=1$ vortices  (Fig.
\ref{ConfinementHalfQFig}). Let us consider the structure of such a
vortex molecule.

Introducing
\begin{equation}
\tilde \Phi=\Phi-{\phi\over 2}~~,~~\tilde \Psi = |\Psi|
e^{i\tilde
\Phi}
\label{NewPhase}
\end{equation}
  one obtains
\begin{equation}
F\{\tilde\Psi\}= {\beta\over 2\rho}\left(|\tilde\Psi|^2-\rho_0\right)^2+
{1\over 2}{\hbar^2\over m^2}\left| \nabla \tilde\Psi\right|^2 + \tilde\alpha
|\tilde\Psi|^2
\sin^2 \tilde\Phi  ~.
\label{FGL2}
\end{equation}
This equation is completely uncoupled from equation for ${\bf v}_{\rm s}$.
However, in the presence of half-quantum vortices there is a topological
connection due to the Aharonov-Bohm effect: for the
Bose condensate, the half-quantum vortex in the pair condensate is 
viewed as the
Aharonov-Bohm tube with the half-quantum magnetic flux.   Since
$\phi$ has
$2\pi$ winding around each half-quantum vortex, the phase 
$\tilde\Phi$ must have
a
$\pi$-jump across some line terminating on a half-quantum vortex. The structure
of the whole system can be easily found in two extreme cases
determined by the Josephson  coupling
$\tilde\alpha$.

If the Josephson
coupling is big, the phase of the Bose-condensate $\Phi$ is strongly pinned by
the phase $\phi$ of the pair condensate, and one has either
$\tilde\Phi=0$ or
$\tilde\Phi=\pi$. In this case the $\pi$-jump is realized due to the
$\pi$-soliton in Fig.
\ref{ConfinementHalfQFig}(a). The Bose-condensate wave function can be
represented as
\begin{eqnarray}
\nonumber{\tilde\Psi(x,y)\over \sqrt{\rho_0}}
= \left(\Theta(-x-x_0)+\Theta(x-x_0)\right){\rm sign}~y~+~\\
+~\Theta(x+x_0)
\Theta(x_0-x)\tanh {y\over \xi} ~,
\label{Soliton}
\end{eqnarray}
where $\Theta$ is the step function; $(x,y)=(x_0,0)$ and $(x,y)=(-x_0,0)$ are
positions of half-quantum vortices, which are topologically confined by the
$\pi$-soliton; and $\xi$ is the coherence length of the Bose-condensate.
The phase $\tilde\Phi$ has $\pi$-jumps on the dashed lines terminating on
half-quantum vortices in Fig. \ref{ConfinementHalfQFig}(a).

If the  tension $\sigma$ of the
$\pi$-soliton is known, the distance $R=2x_0$ between the Alice strings in
such a non-axisymmetric vortex can be easily found from the 
consideration of the
$R$-dependent part of the energy of the vortex per unit length:
\begin{equation}
U_{\rm vortex}(R)=\sigma R - {\pi\over 2}{\hbar^2\over m^2}\rho\ln~R ~.
\label{VortexEnergy}
\end{equation}
Here the first term is the energy of confinement of half-quantum
vortices due to the tension of the
soliton, and the second term is the hydrodynamic repulsion of half-quantum
vortices. Minimization gives the equilibrium distance:
\begin{equation}
   R =  {\pi\hbar^2\rho\over 2\sigma m^2} ~.
\label{VortexDistance}
\end{equation}
For dilute Bose condensate, the coherence length of the Bose 
condensate is large
compared with the coherence length of the pair condensate, $\xi\gg a$; the
coherence length of the pair condensate $a$, which determines the core of the
Alice string, is on the order of interatomic distance. Taking into account that
the relative density of the Bose condensate $\rho_0/\rho \sim a^2/\xi^2$; the
surface tension
$\sigma\sim
\hbar^2/(ma\xi^3)$; and
$\rho\sim m/a^3$, one obtains
$R/a\sim \xi^3/a^3\sim (\rho/\rho_0)^{3/2}\gg 1$.

In the other extreme case, when the pinning of the Bose-condensate phase is
weak, it is more advantageous to fix the order parameter magnitude
$|\Psi|=\sqrt{\rho_0}$, and vary the phase
$\tilde\Phi$. In this configuration instead of
$\pi$-soliton, the Bose condensate contains the fat soliton with
the $N=1$ vortex in the center   (Fig.
\ref{ConfinementHalfQFig}(b)). Within the fat soliton the phase of the
condensate changes from $\pm \pi/2$ to $0$ in the region of thickness
$\tilde\xi=\hbar/(m\tilde\alpha^{1/2})$. The tension of the fat soliton,
which now enters the confinement term in Eq.(\ref{VortexEnergy}), is
$\sigma\sim\tilde\xi \tilde\alpha\rho_0 \sim \tilde\alpha^{1/2}\rho_0
(\hbar/m) \sim
\hbar^2/(ma\xi^2\tilde\xi)$.

The weak-pinning regime occurs when $\tilde\xi\gg
\xi$, while the strong-pinning regime occurs when $\tilde\xi\ll
\xi$. In both regimes, and also in the intermediate regime
when $\tilde\xi\sim\xi$, the structure of the vortex core is highly 
anisotropic,
and the core size considerably exceeds the interatomic distance $a$.

  In conclusion, if the idea that the dominating non-condensate particles form
the pair-correlated state is valid for superfluid $^4$He, vortices in 
superfluid
$^4$He must be highly anisotropic in the limit $\rho_0/\rho\rightarrow 0$.

I thank  N.\,B.\,Kopnin for discussion. This work was supported by ESF COSLAB
Programme and by the Russian Foundations for Fundamental Research.

\end{document}